\documentclass[12pt]{article}
\usepackage[dvips]{color}
\usepackage{epsfig}
\usepackage{amsmath}
\usepackage{graphicx}

\begin{document}
\title{\bf FRW cosmology with the extended Chaplygin gas}
\author{{B. Pourhassan\thanks{Email: bpourhassan@yahoo.com}, E.O. Kahya\thanks{Email: eokahya@itu.edu.tr}}\\
{\small {\em Physics Department, Istanbul Technical University, Istanbul, Turkey}}}  \maketitle
\begin{abstract}
In this paper, we propose extended Chaplygin gas equation of state for which it recovers barotropic fluid with quadratic equation of state. We use numerical method to investigate the behavior of some cosmological parameters such as scale factor, Hubble expansion parameter, energy density and deceleration parameter. We also discuss about the resulting effective equation of state parameter. Using density perturbations we investigate the stability of the theory.
{\bf Pacs Number(s):} 95.35.+d, 95.85.-e, 98.80.-k, 95.30.Cq, 97.20.Vs, 98.80.Cq
\end{abstract}
\section{Introduction}
Accelerated expansion of universe may be described by dark energy which has positive energy and adequate
negative pressure [1, 2]. There are several theories to describe the dark
energy such as quintessence [3]. Another candidate is Einstein's
cosmological constant which has two crucial problems so called fine tuning and
coincidence [4]. There are also
other interesting models to describe the dark energy such as
$k$-essence model [5] and tachyonic model [6]. An interesting
model to describe dark energy is Chaplygin gas [7, 8], that are based on Chaplygin equation (CG) of state [9], which are not good consistent with
observational data [10]. Therefore, an extension of CG model proposed [11-13], which is called generalized Chaplygin gas
(GCG). It is also possible to study viscosity in GCG [14-19]. However, observational data ruled out such a proposal. Then,
GCG was extended to the modified Chaplygin gas (MCG) [20]. Recently, viscous MCG is also suggested and studied [21, 22]. A further extension of CG model is called modified cosmic Chaplygin gas (MCCG) was recently proposed [23-25].
Also, various
Chaplygin gas models were studied from the holography point of view [26-28].\\
The MCG equation of state (EoS) has two parts, the first term gives an ordinary fluid obeying a linear barotropic EoS, and the second term relates pressure to some power of the the inverse of energy density. So here we are essentially dealing with a two-fluid model. However, it is possible to consider barotropic fluid with quadratic EoS or even with higher orders EoS [29, 30]. Therefore, it is interesting to extend MCG EoS which recovers at least barotropic fluid with quadratic EoS.\\
MCG is described by the following EoS,
$$p=A\rho-\frac{B}{\rho^{\alpha}}$$
Now we would like to introduce the extended Chaplygin gas EoS,
$$p=\sum_{n}A_{n}\rho^{n}-\frac{B}{\rho^{\alpha}},$$
which reduces to MCG EoS for $n=1$ ($A_{0}=0$) and can recover barotropic fluid with quadratic EoS by setting $n=2$. Also higher $n$ may recover higher order barotropic fluid which is indeed our motivation to suggest extended Chaplygin gas. We hope this model will be consistent with observational data compared to previous models.\\
This paper organized as follows. In section 2 we review FRW cosmology and give some useful equations to study cosmological parameters. In section 3 we introduce our model and numerically analyze some cosmological parameters. In section 4 we study the deceleration parameter and compare our results with some observational data. In section 5 we investigate the stability of our model and study density perturbations and speed of sound in the same context. Finally in section 6 we summarize our results and give a conclusion.

\section{Equations}
The spatially flat Friedmann-Robertson-Walker (FRW) universe is described by the following metric,
\begin{equation}\label{s1}
ds^2=dt^2-a(t)^2(dr^2+r^{2}d\Omega^{2}),
\end{equation}
where $d\Omega^{2}=d\theta^{2}+\sin^{2}\theta d\phi^{2}$. Also, $a(t)$
represents time-dependent scale factor. The energy-momentum tensor for a perfect fluid is given by,
\begin{equation}\label{s2}
T_{\nu}^{\mu}=(\rho+p)\delta_{0}^{\mu}\delta_{\nu}^{0}-p\delta_{\nu}^{\mu},
\end{equation}
where $\rho(t)$ is the energy density and $p(t)$ is the isotropic pressure. Also, $u^{0}=1$ and $u^{i}=0$ ($i=1, 2, 3$) with $g^{\mu\nu}u_{\mu}u_{\nu}=1$. The independent field equations for the metric (1) and the energy-momentum
tensor (2) are given by,
\begin{equation}\label{s3}
3H^{2}=3(\frac{\dot{a}}{a})^{2}=\rho,
\end{equation}
and,
\begin{equation}\label{s4}
2\frac{\ddot{a}}{a}+(\frac{\dot{a}}{a})^{2}=-p,
\end{equation}
where dot denotes derivative with respect to the cosmic time $t$, and we take $8\pi G = 1$.
It is also assumed that the total matter and energy are
conserved with the following conservation equation,
\begin{equation}\label{s5}
\dot{\rho}+3\frac{\dot{a}}{a}(\rho+p)=0.
\end{equation}

\section{Extended Chaplygin gas EoS}
Modified Chaplygin gas was introduced with the
following equation of state,
\begin{equation}\label{s6}
p=A\rho-\frac{B}{\rho^{\alpha}},
\end{equation}
where $0<A<1/3$, $B$ and $0<\alpha<1$ are positive constants. In this model, one gets a constant negative pressure at low energy
density and high pressure at high energy density. Choosing $A = 0$ one gets generalized Chaplygin gas EoS, and $A = 0$ together $\alpha = 1$ recovers the
original Chaplygin gas EoS. Moreover, the first term on the r.h.s. of the equation (6) gives an
ordinary fluid obeying a barotropic EoS, while there are other barotropic fluids with EoS being quadratic and higher orders [29]. Since modified Chaplygin gas can only recover linear form
of barotropic EoS, here we would like to extend this model so that resulting EoS also can
recover EoS of barotropic fluids  with higher orders. In that case we propose the following EoS,
\begin{equation}\label{s7}
p=\sum_{n}A_{n}\rho^{n}-\frac{B}{\rho^{\alpha}},
\end{equation}
which is called the extended Chaplygin gas EoS. Now, $n=1$ recovers ordinary MCG with $A_{1}=A$. In order to obtain the scale factor-dependence of energy density we should use EoS given by (7) in conservation equation (5). In the following special cases we only consider the last term of expansion in Eq. (7) and find special solution.
\subsection{$n=1$}
Special case of $n=1$ reduces to the modified Chaplygin gas EoS with the following density [31],
\begin{equation}\label{s8}
\rho=\left[\frac{B}{1+A}+\frac{C}{a^{3(1+\alpha)(1+A)}}\right]^{\frac{1}{1+\alpha}},
\end{equation}
where $C$ is an integration constant, and as we mentioned above, the last term of expansion considered. Therefore, we can obtain Hubble parameter as the following,
\begin{equation}\label{s9}
H=\frac{\dot{a}}{a}=\frac{1}{\sqrt{3}}\left[\frac{B}{1+A}+\frac{C}{a^{3(1+\alpha)(1+A)}}\right]^{\frac{1}{2(1+\alpha)}}.
\end{equation}
This case completely discussed in the Ref. [32].
\subsection{$n=-\alpha$}
Here, we assume that the last term of expression in EoS (7) is dominant.
In that case one can express energy density in terms of scale factor as,
\begin{equation}\label{s10}
\rho=\left[B-A+\frac{C}{a^{3(1+\alpha)}}\right]^{\frac{1}{1+\alpha}},
\end{equation}
where $C$ is an integration constant.
\begin{figure}[th]
\begin{center}
\includegraphics[scale=.27]{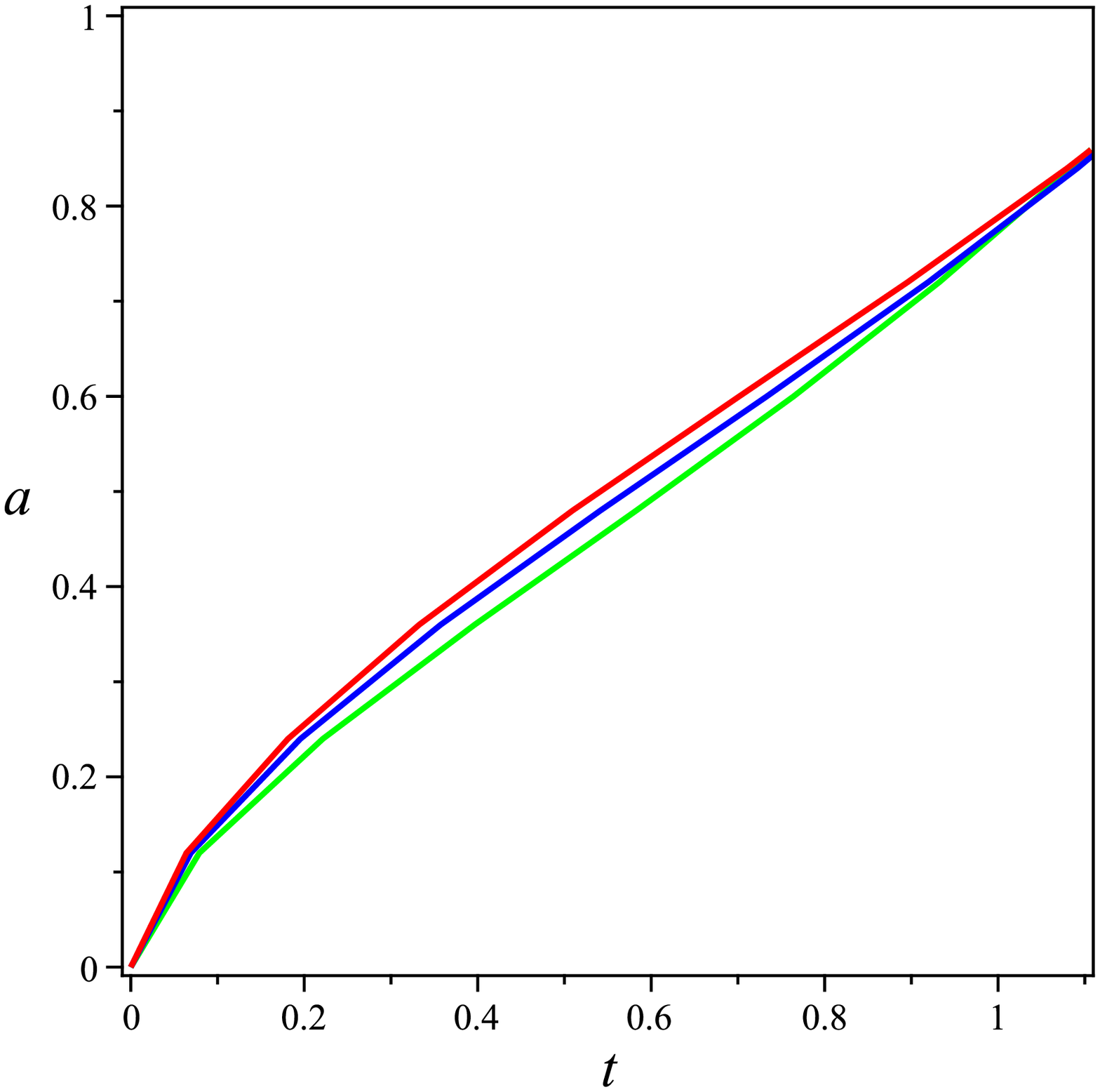}\includegraphics[scale=.27]{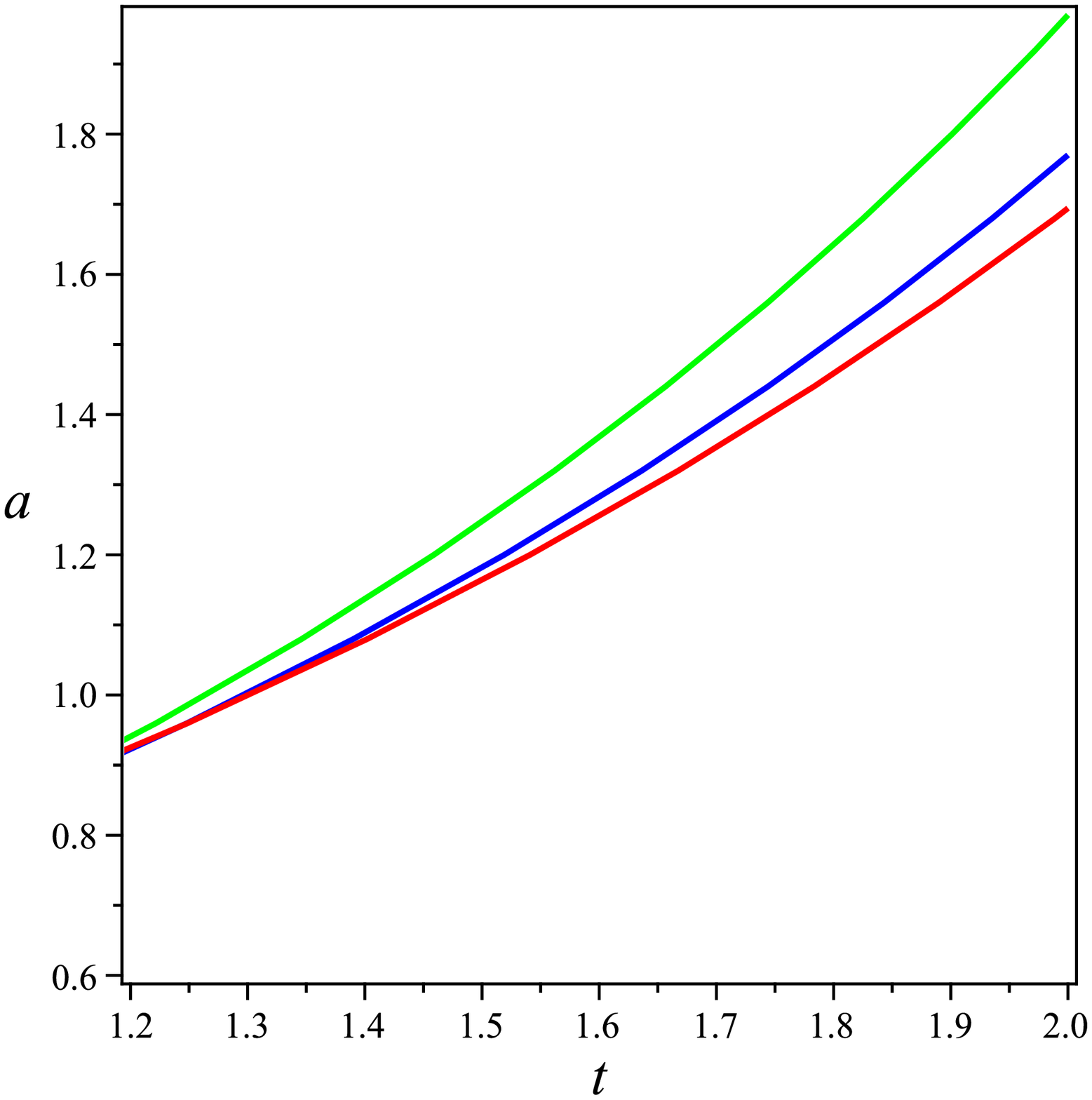}
\caption{Scale factor versus time for $B=3$ and $A=1/3$. $\alpha=0.1$ (green line), $\alpha=0.5$ (blue line), $\alpha=0.9$ (red line).}
\end{center}
\end{figure}

We can use numerical method to solve the Friedmann equations and obtain the time dependence of the scale factor in plots of Fig. 1 and Fig. 2 for various parameters. In Fig. 1 we fix $A$ and $B$, and see that the variation of $\alpha$ has opposite effect at the early and late time. In the early universe, increasing $\alpha$ increases the value of the scale factor (left plot of the Fig. 1). But at the late times, increasing $\alpha$ decreases the value of the scale factor (right plot of the Fig. 1). In the Fig. 2 (left) we fix $A$ and $\alpha$ and vary $B$. We find that, in the early universe the value of $B$ is not important, and it is reasonable because density is high at the initial stage and the second term of EoS becomes negligible. But, at the late time, increasing $B$ increases the value of the scale factor. Another interesting case is when we set $A=\frac{1-\alpha}{1+\alpha}$ and fix $B$ to investigate time-dependent scale factor by variation of $\alpha$ (see right plot of Fig. 2). Under this assumption we get from the equation (7)
that as $a$ increases ($0 < \alpha < 1$), the first term on the r.h.s. reduces to zero while
the second term decreases for a chosen energy density. Special case of $\alpha=0.5$ yields to $A=1/3$ which is illustrated by blue line of the Fig. 2 (right).\\

\begin{figure}[th]
\begin{center}
\includegraphics[scale=.27]{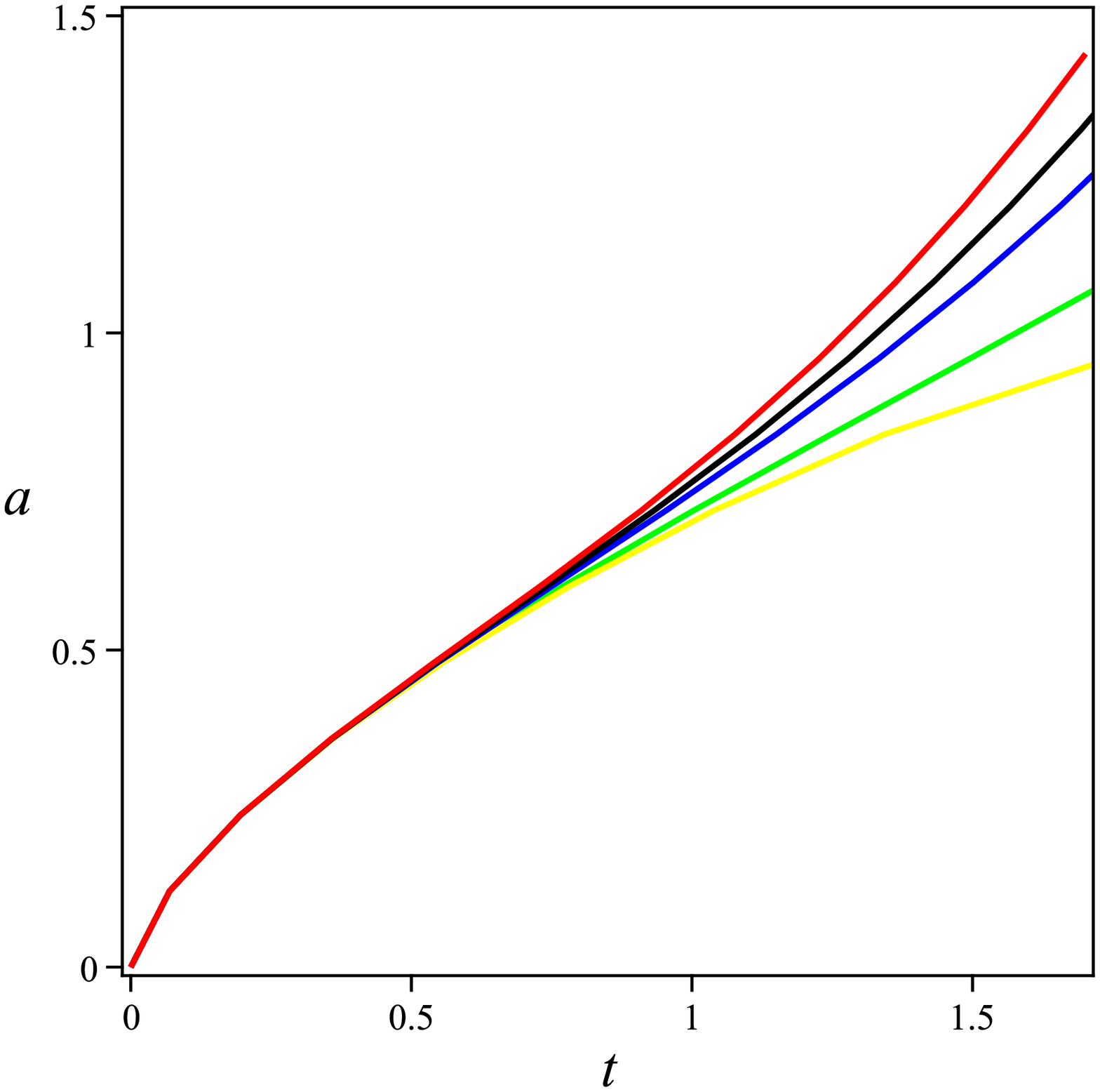}\includegraphics[scale=.27]{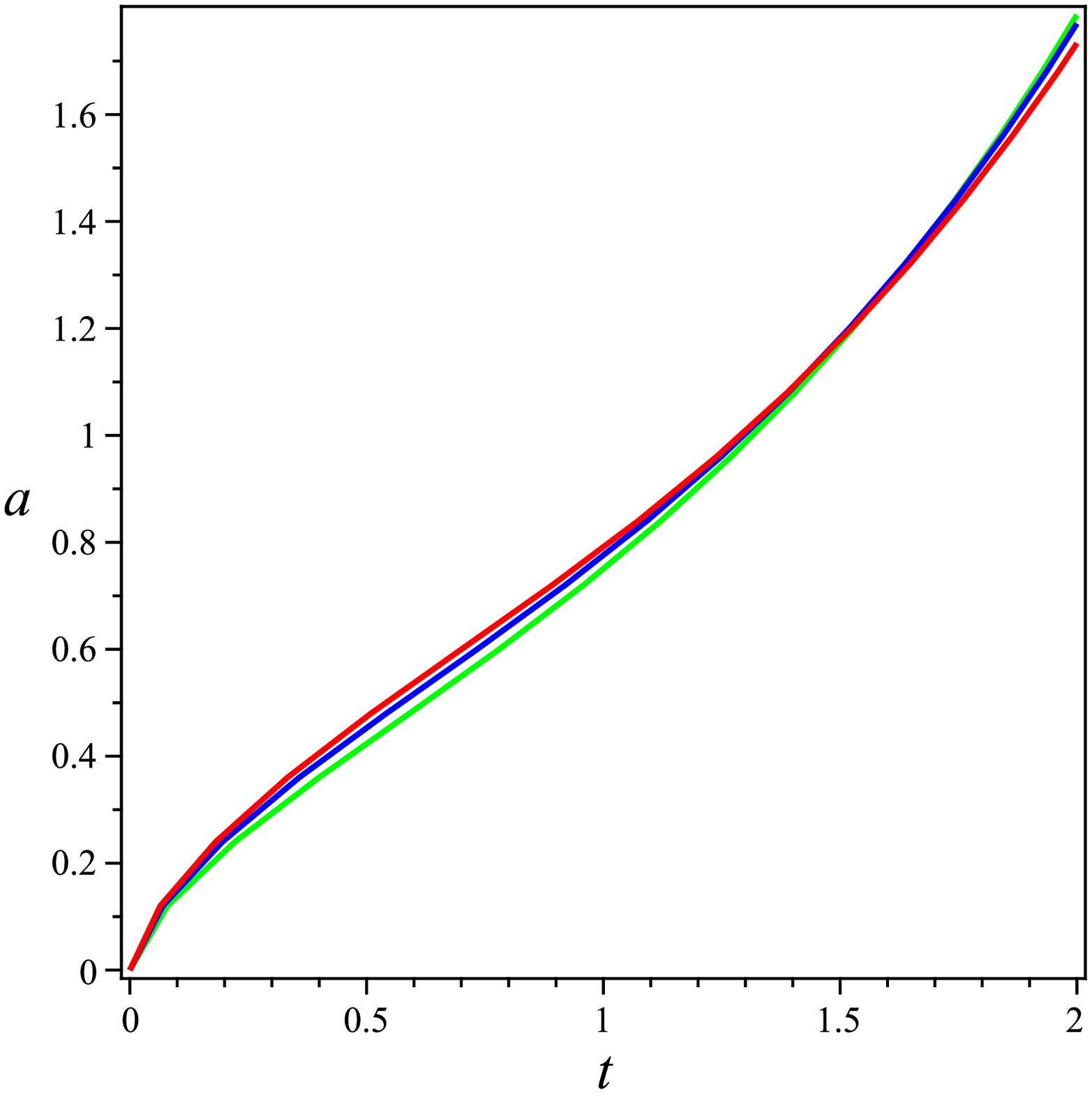}
\caption{Scale factor versus time for; Left:$\alpha=0.5$ and $A=1/3$. $B=0$ (yellow) $B=0.6$ (green), $B=1.8$ (blue), $B=2.5$ (black), $B=3.4$ (red). Right: $B=3$ and $A=\frac{1-\alpha}{1+\alpha}$. $\alpha=0.1$ (green line), $\alpha=0.5$ (blue line), $\alpha=0.9$ (red line).}
\end{center}
\end{figure}

In order to compare this state with observational data we study Hubble expansion parameter in terms of redshift in the Fig. 3 for selected values of parameters $B$, $A$, $\alpha$, and constant $C$. We can see that current value of the Hubble expansion parameter obtained as $H_{0}\sim70$ corresponding to $C=1$ which is near several observational data [33]. Also, we can use observational data given by Refs. [34, 35] to compare $H(Z)$ at different redshifts.  We can see from Fig. 3 that the later value of $H(Z)$ has higher value than results presented by Refs. [34, 35]. In order to have agreement with these data we can choose smallest value of the constant $C$. It is clear that the dashed line of Fig. 3 is near the results obtained by SJVKS10 or best fitted values of [34, 35]. However, we can obtain exact agreement by choosing appropriate small value of $C$.

\begin{figure}[th]
\begin{center}
\includegraphics[scale=.27]{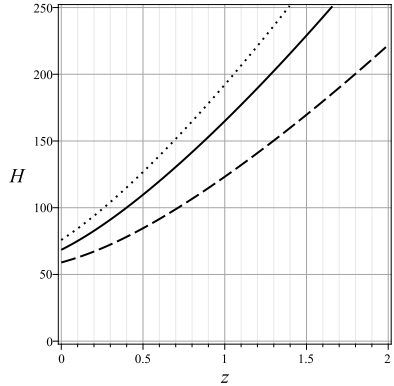}
\caption{Hubble expansion parameter versus redshift for $B=1$, $A=1/3$, and $\alpha=0.9$ for $C=0.4$ (dashed line), $C=1$ (solid line) and $C=1.6$ (dotted line), dots denotes observational data [36].}
\end{center}
\end{figure}

\subsection{$\alpha=-1$}
Similar to the previous case, we focus on the last summation term of the equation (7).
In that case, one can express energy density in terms of the scale factor as,
\begin{equation}\label{s11}
\rho=\left[\frac{A}{B-1}-\frac{C}{a^{3(B-1)(n-1)}}\right]^{\frac{1}{1-n}},
\end{equation}
where $C$ is an integration constant.
Therefore, we can obtain Hubble parameter as the following,
\begin{equation}\label{s12}
H=\frac{\dot{a}}{a}=\frac{1}{\sqrt{3}}\left[\frac{A}{B-1}-\frac{1}{3(B-1)a^{3(B-1)(n-1)}}\right]^{\frac{1}{2(1-n)}}.
\end{equation}
This is indeed dual of the first case (subsection 3.1) by $A\rightarrow-B$ and $n\rightarrow-\alpha$.
\subsection{$n=2$ and $\alpha=1/2$}
In that case the EoS (7) reduces to the following expression,
\begin{equation}\label{s13}
P=A_{1}\rho+A_{2}\rho^{2}-\frac{B}{\sqrt{\rho}}.
\end{equation}
We assume that $A_{1}=A_{2}\equiv A$, and use Eq. (5) to obtain the following integral,
\begin{equation}\label{s14}
\ln(a)=-\int{\frac{d\rho}{3((1+A)\rho+A\rho^{2}-\frac{B}{\sqrt{\rho}})}}.
\end{equation}
It gives us the following energy density,
\begin{equation}\label{s15}
\rho=\frac{(Xa^{\frac{9}{2}}+a^{-\frac{9+15X^{2}}{2}A})^{2}}{a^{9}},
\end{equation}
where $X$ is root of the following equation,
\begin{equation}\label{s16}
AX^{5}+(1+A)X^{3}-B=0.
\end{equation}
In this case the Hubble parameter in terms of redshift plotted in the Fig. 4, which shows that, this case is far from observations in any time.

\begin{figure}[th]
\begin{center}
\includegraphics[scale=.27]{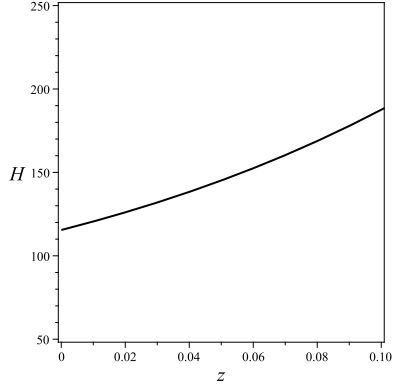}
\caption{Hubble expansion parameter versus redshift in case of $n=2$, for $A=1/3$, and $\alpha=0.5$ for $C=1$.}
\end{center}
\end{figure}

\subsection{$n=3$ and $\alpha=1/2$}
In that case the EoS (7) reduces to the following expression,
\begin{equation}\label{s17}
P=A_{1}\rho+A_{2}\rho^{2}+A_{3}\rho^{3}-\frac{B}{\sqrt{\rho}}.
\end{equation}
We assume that $A_{1}=A_{2}=A_{3}\equiv A$, and use Eq. (5) to find the following integral,
\begin{equation}\label{s18}
\ln(a)=-\int{\frac{d\rho}{3((1+A)\rho+A\rho^{2}+A\rho^{3}-\frac{B}{\sqrt{\rho}})}}.
\end{equation}
It gives us the following energy density,
\begin{equation}\label{s19}
\rho=\frac{(Ya^{\frac{9}{2}}+a^{-\frac{9+15Y^{2}+21Y^{4}}{2}A})^{2}}{a^{9}},
\end{equation}
where $Y$ is root of the following equation,
\begin{equation}\label{s20}
AY^{7}+AY^{5}+(1+A)Y^{3}-B=0.
\end{equation}
Resulting Hubble parameter of this case is similar to the previous case ($n=2$), therefore we check the next case.

\subsection{Arbitrary $n$ and $\alpha$}
In the previous subsections we discussed some particular cases of $n$ and $\alpha$. Now, we would like to consider general case and give numerical analysis of the cosmological parameters such as scale factor, dark energy density and Hubble expansion parameter with an arbitrary choice of $n$ and $\alpha$.
Before doing this, we obtain an expression for the energy density corresponding to $\alpha=0.5$ which is extension of the previous subsections. Repeating procedure of the subsections 3.4 and 3.5 gives us the following expressions,
\begin{equation}\label{s21}
\rho=\frac{(\mathcal{C}a^{\frac{9}{2}}+a^{-\frac{9+\sum_{n}6(n-1){\mathcal{C}}^{2(n-1)}}{2}A})^{2}}{a^{9}},
\end{equation}
where $\mathcal{C}$ is root of the following equation,
\begin{equation}\label{s22}
A\sum_{n}{\mathcal{C}}^{2n+1}+(1+A){\mathcal{C}}^{3}-B=0.
\end{equation}
We can investigate $H(z)$ using the expression,
\begin{equation}\label{s23}
\Omega = \frac{\rho}{3H_0^2},
\end{equation}
where $H_0$ being the Hubble parameter today, in the form,
\begin{equation}\label{s24}
\Omega(z) = \Omega_0\biggr[1 - c + c(1 + z)^r\biggl]^2,
\end{equation}
where $c$ and $r$ are constants related to the constants $\mathcal{C}$, $A$ and $B$. Since there is only one fluid, and the spatial section is flat, $\Omega_0 = 1$. Using the Friedmann's equation, it is possible to obtain $H(z)$ and compare it with the observational data. We perform it in Fig. 5. We can see that choosing $c=0.5$ and $r=1.7$ are the best fit in agreement with observational data [34, 35]. Therefore, always we can choose appropriate values of $c$ and $r$ to have a model in agreement with observational data better than $\Lambda$CDM model.

\begin{figure}[th]
\begin{center}
\includegraphics[scale=.27]{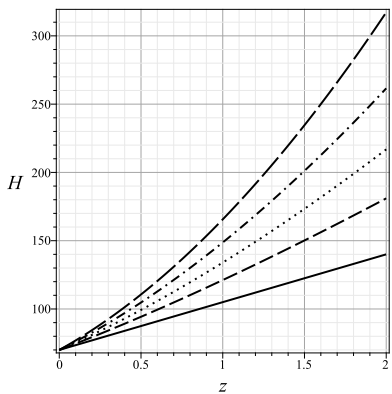}
\caption{Hubble expansion parameter versus redshift in case of arbitrary $n$ and $\alpha$. We fix $c=0.5$ and $r=1$ (solid line), $r=1.3$ (dashed line), $r=1.5$ (dotted line), $r=1.7$ (dash dotted line), $r=1.9$ (long dashed line), dots denotes observational data [36].}
\end{center}
\end{figure}

However, some disagreement of these cases with observational data of $H(z)$ may because of choosing $\alpha=0.5$. Other choice of $\alpha$ should investigate numerically as the following.\\
In order to find real solutions which will be interesting from observational point of view, first of all, we combine equations (3), (4) and (7) to obtain the following second order differential equation,
\begin{equation}\label{s25}
2\frac{\ddot{a}}{a}+(\frac{\dot{a}}{a})^{2}+\sum_{n}3^{n}A(\frac{\dot{a}}{a})^{2n}-3^{-\alpha}B(\frac{\dot{a}}{a})^{-2\alpha}=0,
\end{equation}
where we assume $A_{n}\equiv A$ for simplicity and reducing free parameters. Numerically, we can solve equation (13) and obtain behavior of scale factor against $t$. In Fig. 6 we fix $B$, $\alpha$ and $A$, and vary $n$ to find that increasing $n$ decreases value of the scale factor. The left plot of Fig. 6 shows long term behavior of the scale factor, while the right one shows variation of the scale factor at the early universe. Also, Fig. 7 shows that $a$ decreases by increasing $\alpha$ which is in agreement with the late time behavior of the Fig. 1.\\
On the other hand we can combine equations (3), (5) and (7) to obtain the following first order differential equation of the energy density,
\begin{equation}\label{s26}
\dot{\rho}+\sqrt{3}(\rho^{\frac{3}{2}}+A\sum_{n}\rho^{n+\frac{1}{2}}-B\rho^{-\alpha+\frac{1}{2}})=0.
\end{equation}

\begin{figure}[th]
\begin{center}
\includegraphics[scale=.27]{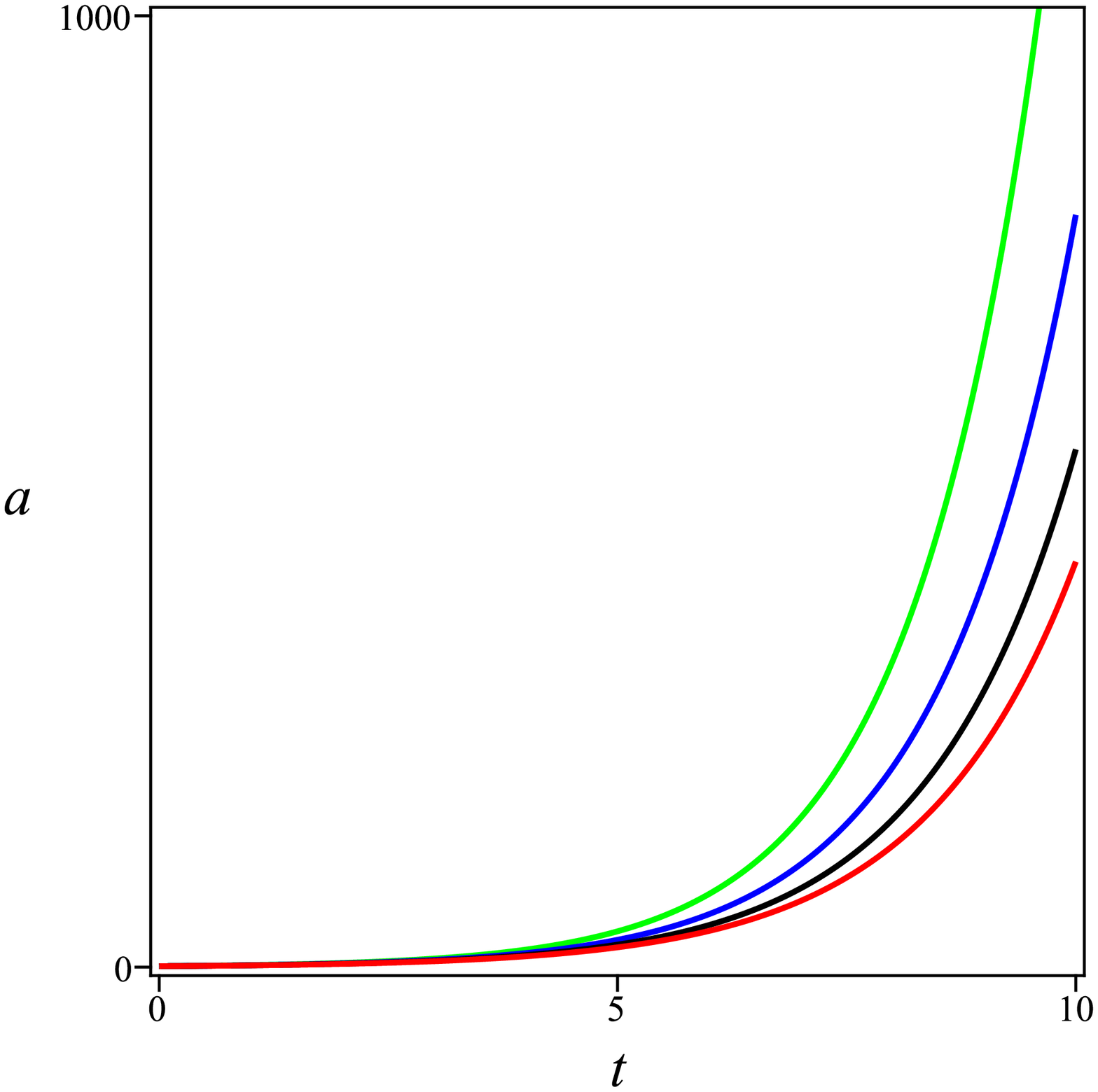}\includegraphics[scale=.27]{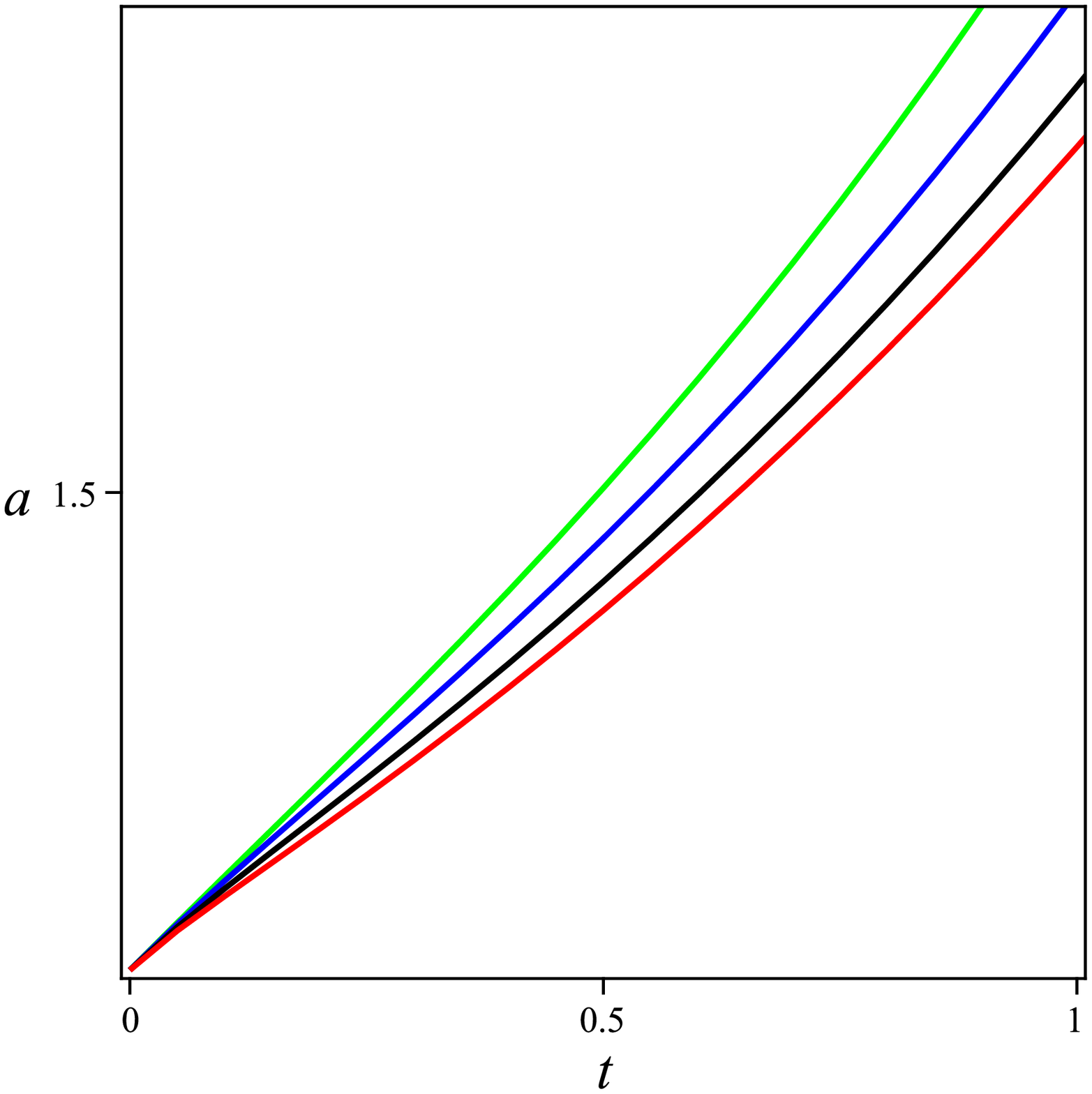}
\caption{Scale factor versus time for $B=3$,  $\alpha=0.9$ and $A=1/3$. $n=1$ (green line), $n=2$ (blue line), $n=3$ (black line), $n=4$ (red line). Left: general behavior. Right: early universe}
\end{center}
\end{figure}

\begin{figure}[th]
\begin{center}
\includegraphics[scale=.27]{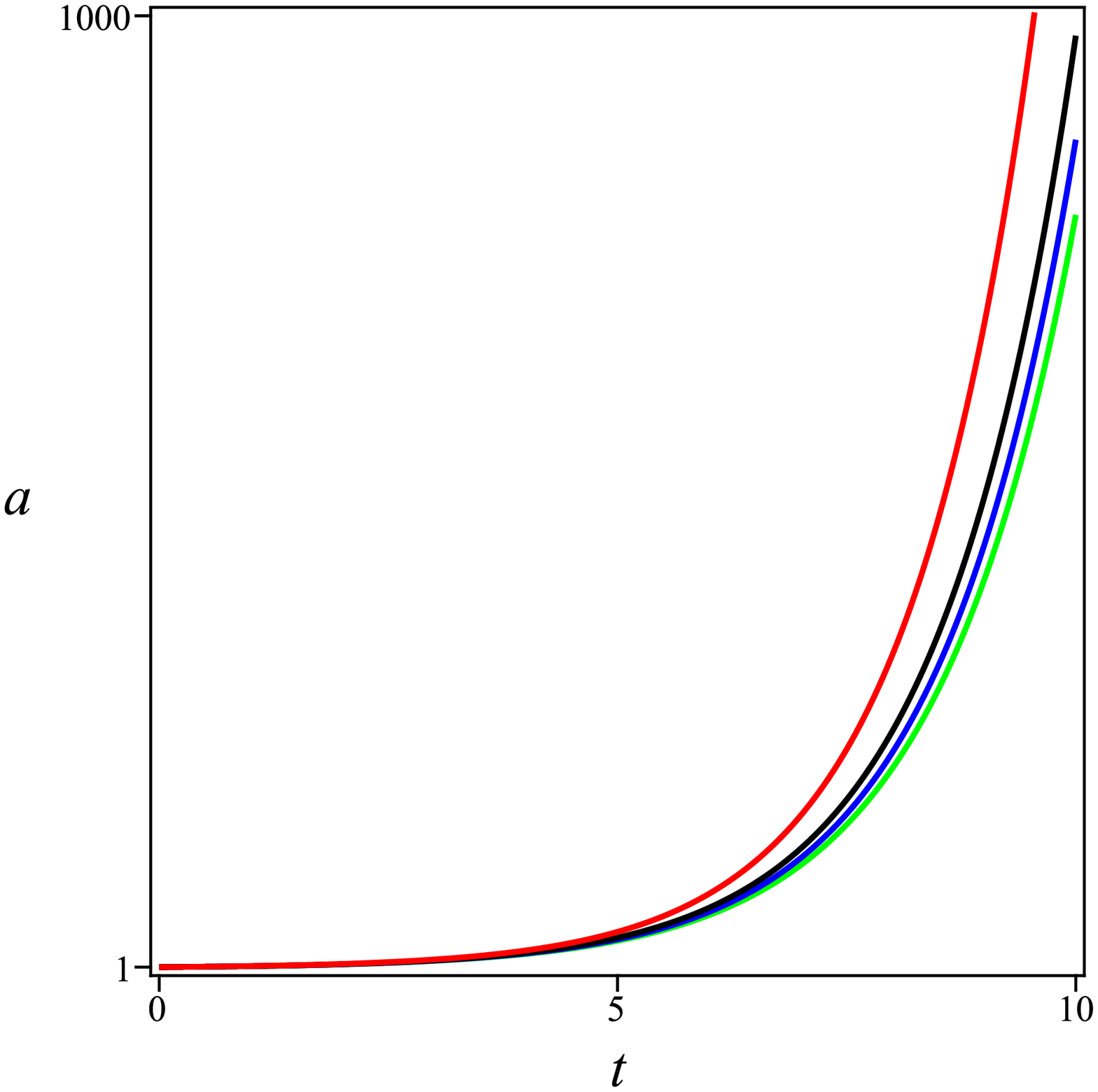}
\caption{Scale factor versus time for $B=3$,  $n=2$ and $A=1/3$. $\alpha=0.9$ (green line), $\alpha=0.7$ (blue line), $\alpha=0.5$ (black line), $\alpha=0.1$ (red line).}
\end{center}
\end{figure}

\begin{figure}[th]
\begin{center}
\includegraphics[scale=.27]{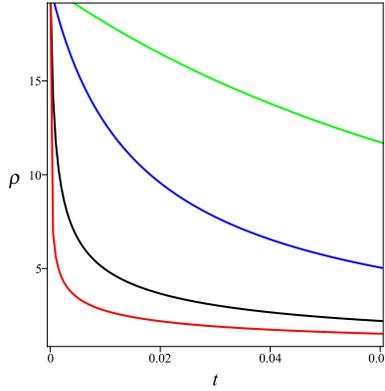}
\caption{Energy density versus time for $B=3$,  $\alpha=0.9$ and $A=1/3$. $n=1$ (green line), $n=2$ (blue line), $n=3$ (black line), $n=4$ (red line).}
\end{center}
\end{figure}

\begin{figure}[th]
\begin{center}
\includegraphics[scale=.27]{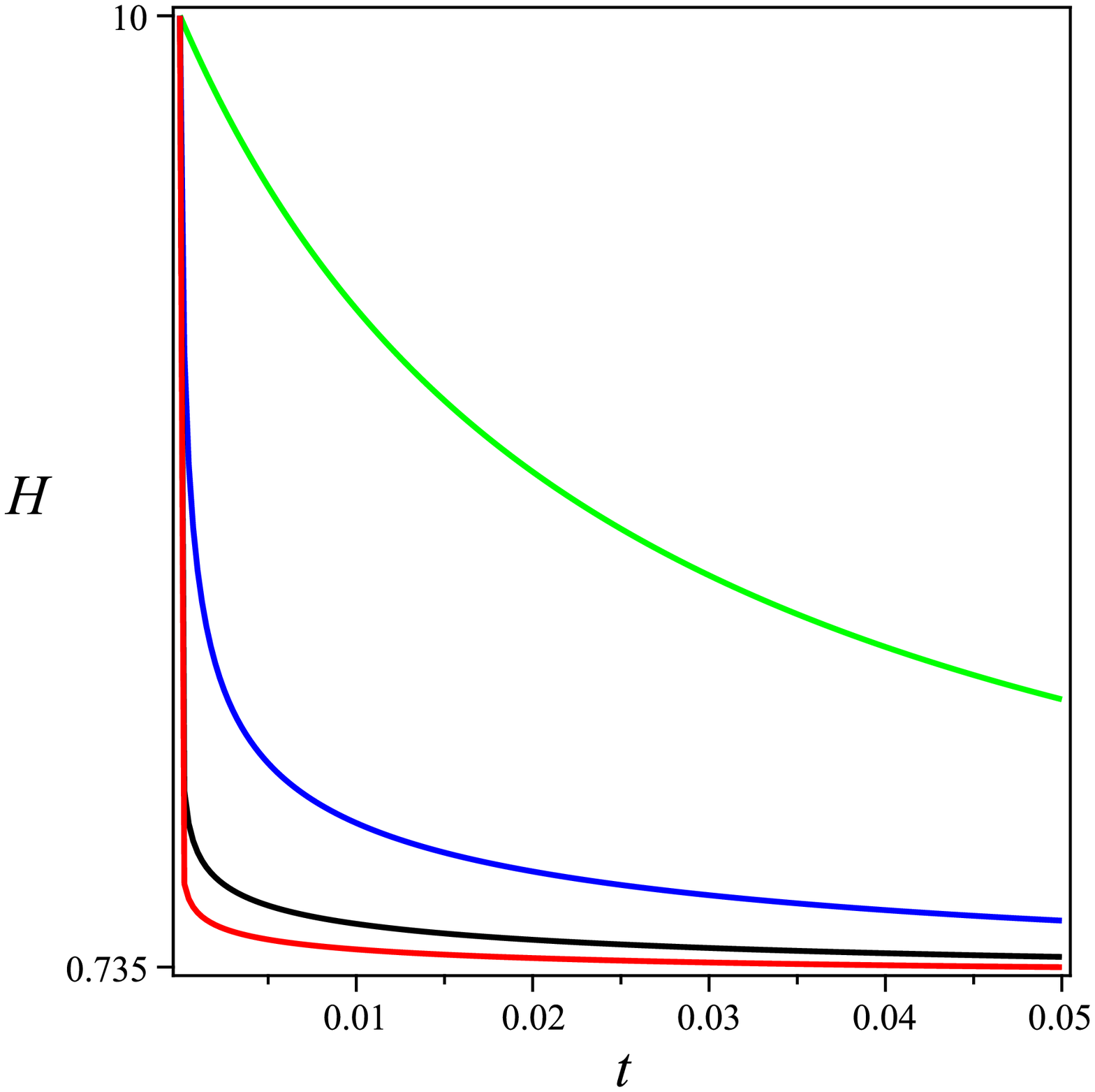}
\caption{Hubble expansion parameter versus time for $B=3$,  $\alpha=0.9$ and $A=1/3$. $n=1$ (green line), $n=2$ (blue line), $n=3$ (black line), $n=4$ (red line).}
\end{center}
\end{figure}
The equation (26) can also be solved numerically to obtain the behavior of the dark energy density. Fig. 8 shows that increasing $n$ decreases value of energy density. As expected, energy density obtained  here is a decreasing function of time which yields to an infinitesimal constant at the late times.\\
By using the equation (3) we can rewrite the equation (4) as follow,
\begin{equation}\label{s27}
2\dot{H}+3H^{2}=-p.
\end{equation}
Now, we use equations (3) and (7) in the equation (27) to study behavior of Hubble expansion parameter via the following equation,
\begin{equation}\label{s28}
2\dot{H}+3H^{2}+\sum_{n}3^{n}AH^{2n}-3^{-\alpha}BH^{-2\alpha}=0.
\end{equation}
Numerical analysis of this equation illustrated in Fig. 9, and shows that increasing $n$ decreases value of the Hubble expansion parameter.

\section{Deceleration parameter}
In the previous section we gave a numerical analysis the behavior of the scale factor, energy density and Hubble expansion parameter. Since it is not clear to see the analytical behavior of these parameters from equations (25), (26) and (28), we would like to look at another parameter to get more information about the  dynamics.\\
An important parameter in cosmology, from theoretical and observational point of views, is called the deceleration parameter which is given by,
\begin{equation}\label{s29}
q=-\left(\frac{\dot{a}}{a}\right)^{-2}\frac{\ddot{a}}{a}=-1-\frac{\dot{H}}{H^{2}}.
\end{equation}
Using equations (3), (7) and (29) one can obtain,
\begin{equation}\label{s30}
q=\frac{1}{2}+\frac{3}{2}\left(\sum_{n}A\rho^{n-1}-\frac{B}{\rho^{\alpha+1}}\right).
\end{equation}
In the Fig. 10 we draw deceleration parameter in terms of $\rho$ for various values of $n$. We can see that increasing $n$ increases the value of $q$. The green line of Fig. 10 is corresponds to modified Chaplygin gas. At the early universe with high density the deceleration parameter may be reduced to the following expression,
\begin{equation}\label{s31}
q\approx\frac{1+3A\sum_{n}\rho^{n-1}}{2}.
\end{equation}
In the case of $n=1$ and $A=-1$ we recover result of $\Lambda$CDM model where $q=-1$. On the other hand, late time behavior (low density limit) of deceleration parameter may be described by,
\begin{equation}\label{s32}
q\approx\frac{1-3B\rho^{-\alpha-1}}{2}.
\end{equation}
Again, special case of $\alpha=-1$ and $B=1$ give $q=-1$.

\begin{figure}[th]
\begin{center}
\includegraphics[scale=.27]{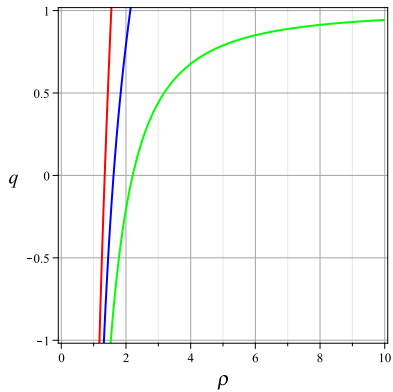}
\caption{Deceleration parameter versus density for $B=3$,  $\alpha=0.9$ and $A=1/3$. $n=1$ (green line), $n=2$ (blue line), $n=3$ (red line).}
\end{center}
\end{figure}

At the late stage of evolution one can obtain an effective EoS parameter as follow,
\begin{equation}\label{s33}
\omega_{eff}=-1+\frac{A}{B}\sum_{n}\rho^{n+\alpha},
\end{equation}
as $\rho\rightarrow0$ then, $\omega_{eff}\rightarrow-1$ so we asymptotically get
$p=-\rho$ from extended Chaplygin gas as well as MCG, which corresponds to an empty universe
with cosmological constant. In Fig. 11 we draw the effective EoS parameter and find that at the late stage, increasing $n$ decreases the value of $\omega_{eff}$ and yields it to -1. On the other hand, at the early universe with high density we have positive effective EoS. It is interesting to note that $\omega_{eff}$ always remains greater than -1, thus avoiding the undesirable feature of big rip similar to the previous cases of Chaplygin gas EoS. Also, we can see that evolution of $\omega_{eff}$ with higher $n$ is faster than the case with lower $n$.

\begin{figure}[th]
\begin{center}
\includegraphics[scale=.27]{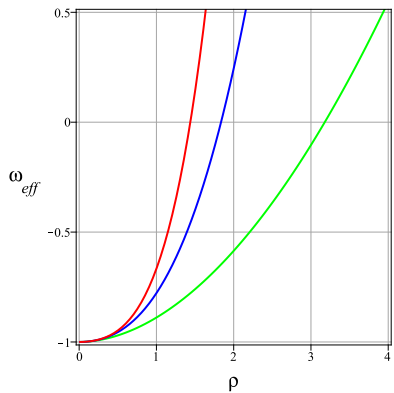}
\caption{Effective EoS parameter versus density for $B=3$,  $\alpha=0.9$ and $A=1/3$. $n=1$ (green line), $n=2$ (blue line), $n=3$ (red line).}
\end{center}
\end{figure}

\section{Density perturbation}
In this section we give density perturbation analysis of our model. For simplicity we neglect the second term in r.h.s of the EoS (7) to investigate the effect of $n$. Already, density perturbation of a universe dominated by Chaplygin gas was studied by the Ref. [33]. Now, we use their results to write the following perturbation equation corresponding to flat FRW universe which expands with acceleration. The perturbation equation of density is given by,
\begin{equation}\label{s34}
\ddot{\delta}+H[2-3(2\omega-C_{s}^{2})]\dot{\delta}-\frac{3}{2}H^{2}(1-6C_{s}^{2}-3\omega^{2}+8\omega)\delta=-k^{2}\frac{C_{s}^{2}}{a^{2}}\delta,
\end{equation}
where $\delta$ is a density fluctuation, $k$ is the wave-number of the
Fourier mode of the perturbation, $\omega=p/\rho$, and,
\begin{equation}\label{s35}
C_{s}^{2}=\frac{\dot{p}}{\dot{\rho}},
\end{equation}
is sound speed. This is an important parameter to investigate stability of the theory. Extended Chaplygin gas with real positive sound speed is stable therefore we should seek regions in which the squared sound speed will be positive to have stability. In the following subsections we solve the equation (34) numerically for some values of $n$.
\subsection{$n=1$}
In the case of $n=1$ one can obtain,
\begin{equation}\label{s36}
\rho=\left[\frac{\sqrt{3}}{2}(1+A)t+C_{1}\right]^{-2},
\end{equation}
where $C_{1}$ is an integration constant. Then, the behavior of $\delta$ is illustrated in Fig. 12, which shows evolution of perturbation for various values of $k$. We can see that at the initial time there is no difference between various values of $k$. After that, increasing $k$ increases the value of $\delta$. Analytical study of a similar case with $\alpha=0.5$ can found in the Ref. [33]. Ref. [33] suggests $\delta$ to be proportional to combination of hypergeometric and exponential function of time which is decreasing function of time at initial stage. Therefore, our results are agree with the Ref. [33]. However, there is also a numerical analysis with $\alpha>1$ which suggest $\delta$ is increasing function of scale factor [38], so this case is not relevant to our study. The $n=1$ case corresponding to MCG, was ruled out by the Ref. [38] by using observational constraints. Therefore, we should consider other cases with higher $n$ to investigate the validity of our model.\\
In the Fig. 13 we fix $C_{1}=0.01$ and study variation of squared sound speed for various values of $\alpha$. We can see that this model is completely stable for $0\leq\alpha\leq1$. However, for larger values of $\alpha$ than 0.2 the causality violated and we lead to sound speed higher than light speed ($c=1$). Therefore, we should use small values of $\alpha$ to obtain appropriate value of sound speed.\\

\begin{figure}[th]
\begin{center}
\includegraphics[scale=.27]{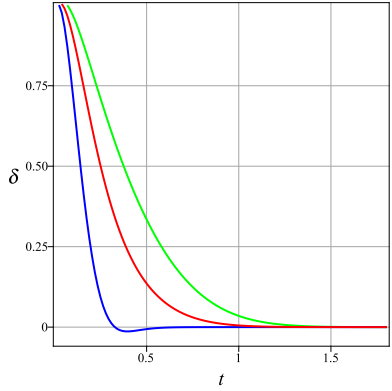}
\caption{Time evolution of $\delta$ for $B=3$, $C_{1}=1$, $n=1$, $A=1/3$ and $\alpha=0.5$. $k=0$ (blue line), $k=5$ (red line), $k=10$ (green line).}
\end{center}
\end{figure}

\begin{figure}[th]
\begin{center}
\includegraphics[scale=.27]{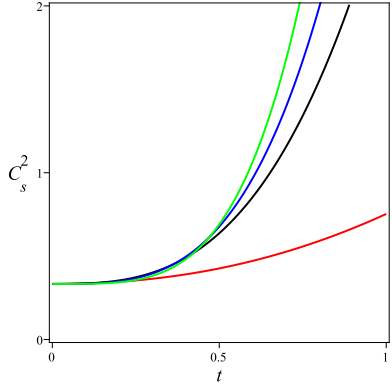}
\caption{Squared sound speed for $B=3$, $C_{1}=0.01$, $n=1$ and $A=1/3$. $\alpha=1$ (green line), $\alpha=0.7$ (blue line), $\alpha=0.5$ (black line), $\alpha=0.1$ (red line).}
\end{center}
\end{figure}

\subsection{$n=2$}
In the case of $n=2$ we can obtain the following dark energy density,
\begin{equation}\label{s37}
\rho=\frac{1+A}{A}\tan^{2}\left(\frac{3(1+A)^{3/2}(t+C_{2})}{2\sqrt{3}A}\right),
\end{equation}
where $C_{2}$ is an integration constant. Therefore, equation (34) can be solved numerically which is illustrated in Fig. 14. This shows time evolution of $\delta$ for various values of $k$. We find that small values of $k$ yield to positive $\delta$ which are decreasing function of time. Larger values of $k$ yields to periodic like $\delta$, which are damped at the late time.\\
Also, in the Fig. 15 we can see the behavior of squared sound speed for some values of $\alpha$. Significantly, we can see a periodic behavior of $C_{s}^{2}$ and find that our model is completely stable.\\
Interesting point in this case is that one can get a stable universe for some non-zero values of $A$ unlike the MCG models that were studied in the Ref. [38]. It gives us good motivation to continue our work to construct a valuable model of the universe.

\begin{figure}[th]
\begin{center}
\includegraphics[scale=.27]{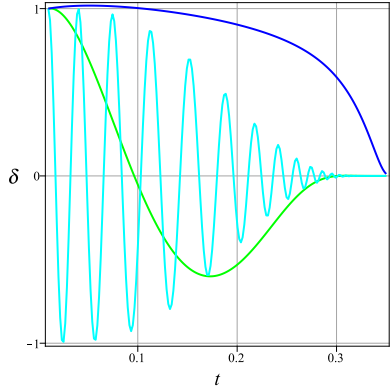}
\caption{Time evolution of $\delta$ for $B=3$, $C_{2}=1$, $\alpha=0.5$, $n=2$ and $A=1/3$. $k=0$ (blue line), $k=10$ (green line), $k=100$ (cyan line).}
\end{center}
\end{figure}

\begin{figure}[th]
\begin{center}
\includegraphics[scale=.27]{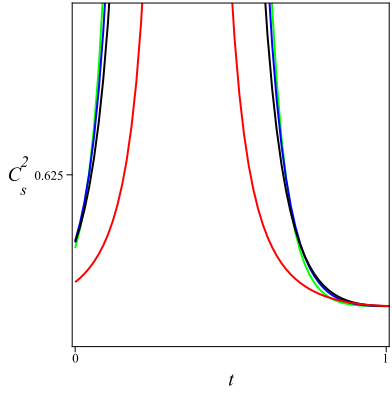}
\caption{Squared sound speed for $B=3$, $C_{2}=1$, $n=2$ and $A=1/3$. $\alpha=1$ (green line), $\alpha=0.7$ (blue line), $\alpha=0.5$ (black line), $\alpha=0.1$ (red line).}
\end{center}
\end{figure}

\subsection{$n=3$}
In that case we can obtain the following time-dependent density,
\begin{equation}\label{s38}
\rho=\frac{1+f(t)+\sqrt{1+2f(t)}}{\sqrt{A}f(t)},
\end{equation}
where,
\begin{equation}\label{s39}
f(t)=\tan^{2}\left(\frac{\sqrt{6}(t+C_{3})}{A^{1/4}}\right),
\end{equation}
and $C_{3}$ is an integration constant. Also, we assumed that the last term of expansion in the equation (7) is dominant.\\
Fig. 16 shows that, $\delta$ grows periodically at the initial time and then behaves as damping periodic function of time. We study evolution of $\delta$ only for $k=0$.\\
Also, Fig. 17 shows a variation of squared sound speed with time which is positive for all values of $\alpha$ and tells that our model is completely stable with $n=3$. Also, there are critical times where the sound speed vanishes and again grows to high value. It means that small value of $\alpha$ should choose to avoid causality.

\begin{figure}[th]
\begin{center}
\includegraphics[scale=.27]{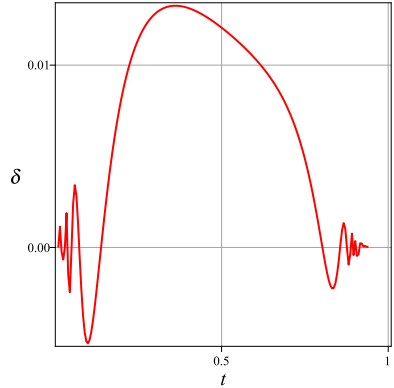}
\caption{Time evolution of $\delta$ for $B=3$, $C_{3}=1$, $\alpha=0.5$, $k=0$ $n=3$ and $A=1/3$.}
\end{center}
\end{figure}

\begin{figure}[th]
\begin{center}
\includegraphics[scale=.27]{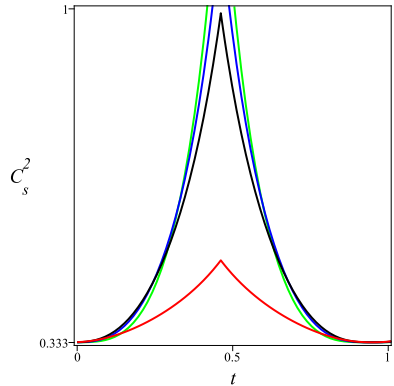}
\caption{Squared sound speed for $B=3$, $C_{3}=1$,  $n=3$ and $A=1/3$. $\alpha=1$ (green line), $\alpha=0.7$ (blue line), $\alpha=0.5$ (black line), $\alpha=0.1$ (red line).}
\end{center}
\end{figure}

Before end of this section, it may be useful to present sound speed in terms of scale factor. In that case we recall special cases of $n=2$ and $n=3$ with $\alpha=0.5$, which is discussed in the subsections 3.4 and 3.5. In that case, Fig. 18 shows that the squared sound speed is positive for both cases. In both cases of $n=2$ and $n=3$ the sound speed yields to a constant for the large scale factor. It is clear that, increasing $n$ increases sound speed. We can fix parameter to have well defined sound speed.\\

\begin{figure}[th]
\begin{center}
\includegraphics[scale=.27]{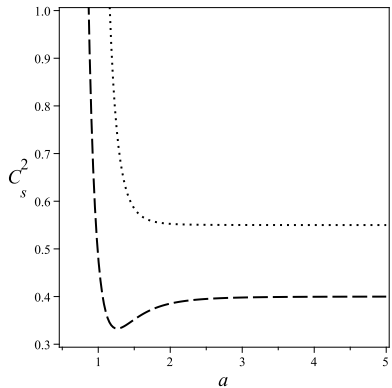}
\caption{Squared sound speed for $B=0.5$, $\alpha=0.5$ and $A=0.05$. $n=2$ (dashed line), $n=3$ (dotted line).}
\end{center}
\end{figure}

\section{Conclusion}
In this paper, an extended model of Chaplygin gas (ECG) as a model of dark energy is proposed which recovers barotropic fluid with quadratic EoS. Scale-factor dependence energy density is obtained for special cases. In the general case, we obtained the evolution of scale factor, Hubble expansion parameter and time-dependent dark energy density, and found the effect of $n$ in cosmological parameters. For instance we found that evolution of scale factor corresponding to $n=1$ (linear barotropic fluid) is faster than the case with $n=2$ (quadratic barotropic fluid). We also found that Hubble expansion parameter and dark energy density are decreasing with $n$.\\
Then, we investigated deceleration parameter and discussed about initial time and late time behavior of it. We have shown that $q\rightarrow-1$ is verified for low densities (late time). Then, we discussed about effective EoS parameter and confirmed that $\omega\geq-1$ is valid also in our model.\\
We analyzed $H(z)$ and compare our results with observational data. We found that, by choosing appropriate values of constant parameters, our model has more agreement with observational data than $\Lambda$CDM.\\
Finally, we studied density perturbations and investigated the stability of our model under assumption that the first term of the equation (7) was dominant. We focused on the special case of $n=1$ (MCG), $n=2$ and $n=3$. We found that the cases of $n=2$ and $n=3$ are completely stable by choosing appropriate values of parameters.\\
We found that adding higher order terms which recovers second and higher order barotropic EoS, also may solve the problem of MCG which was ruled out [38]. Therefore, we concluded that ECG may be a more appropriate model than MCG and GCG and has agreement with the observational data.
This paper is one of the first steps to introduce extended Chaplygin gas model and there are many things to investigate in future works such as construction of holographic version of this model.\\
An important point is that our solution not recover the standard cosmological model in the past, since the extended Chaplygin gas does not behaves as a dust component. In order to have a model which behaves as a dust we should consider varying $A_{n}(t)$ in the equation (7), so at the initial time $A_{n}(t)\rightarrow0$ and gives $p=0$. This is also left for future work.

\end{document}